# Fundamentally different strategies of gene regulation revealed by analysis of binding motifs

**Zeba Wunderlich[1] and Leonid A. Mirny[2]**

[1] Department of Systems Biology, Harvard Medical School, Boston, MA 02155
[2] Corresponding author. Harvard-MIT Division of Health Sciences and Technology, Massachusetts Institute of Technology, Cambridge, MA 02139, E-mail: leonid@mit.edu, Tel: +1-617-452-4862

**Running title:** Information theory and gene regulation.

**Abstract:**
Coordinated regulation of gene expression in the cell relies on molecular processes of transcription factors (TF) binding specific DNA sites. Classical experiments in bacteria demonstrated that TF-promoter binding is sufficient to regulate gene expression. Growing experimental evidences in eukaryotes, however, challenge this association showing little correlation between gene expression and TF binding, with later exhibiting widespread non-functional binding to DNA. This discrepancy can be explained by either limitations of available whole-genome measurements or by a fundamental uncoupling between TF binding and gene expression in eukaryotes. Our study provides strong support for the later hypothesis. Novel information-theoretical analysis of 319 known TF binding motifs clearly demonstrates that prokaryotes and eukaryotes use strikingly different strategies to target TFs to specific genome locations: while bacterial TFs recognize DNA specifically enough to bind a single genomic site, eukaryotic TFs exhibit widespread nonfunctional binding and require clustering of sites in a regulatory region to achieve specificity. Supported by comparative genomics and a wide range of experimental evidences proposed molecular mechanism provides a new framework for interpretation of functional genomic measurements.

**Character count:** 18,538 characters (including spaces)

## Introduction

The binding of a TF to its cognate site is believed to be necessary and sufficient to trigger a cascade of events that regulate gene expression. This has been proved definitively in bacteria, and it has been assumed to be true in eukaryotes. However, growing experimental data on the *in vivo* binding of eukaryotic TFs demonstrates widespread nonfunctional binding of TFs(Li *et al*, 2008) and the lack of correlation between TF binding and differential gene expression(Hu *et al*, 2007), suggesting that TF binding and its functional consequences are uncoupled in eukaryotes. Here we propose that this uncoupling is enabled by the fundamentally different machinery used by eukaryotic TFs to recognize their sites. Strikingly, this uncoupling between binding and gene regulation in eukaryotes is evident from the motifs of TFs. By examining a collection of TF binding motifs, we demonstrate that prokaryotes and eukaryotes use markedly different strategies to "address" a particular location in the genome. The high specificity of prokaryotic TFs targets them precisely to functional cognate sites, while the low specificity of eukaryotic TFs allows non-functional binding, but also provides an opportunity for combinatorial regulation.

## Results and Discussion

We use tools of information theory (Stormo and Fields, 1998) to characterize known motifs of bacterial and eukaryotic TFs. In a genome of size $N$ bps, a minimum of $I_{\min} = \log_2(N)$ bits of information is needed to specify a unique address in that genome. We can compare this value to the information content ($I$) of actual TF binding motifs using the Kullback-Leibler (KL) distance between the motif and the overall genome composition(Schneider *et al*, 1986):

$$I = \sum_{i=1}^{L} \sum_{b \in \{A,C,G,T\}} p_i(b) \log_2 \left( p_i(b)/q(b) \right)$$

Here, $L$ is the length of the motif, $p_i(b)$ is the frequency of base $b$ at position $i$ in the motif, and $q(b)$ is its background frequency. If $I < I_{\min}$, the motif does not have enough information to specify a unique address in a random genome of size $N$. Insufficient information has two consequences: (i) a large number of spurious matches of the motif are present in the genome, creating the potential for widespread nonfunctional binding; and (ii) several sites are required to specify a regulatory region.

We calculated these quantities for 319 prokaryotic and eukaryotic TFs and found that they were markedly different (Figure 1). The average information content of a prokaryotic motif, 19.8 bits, is close to the required $I_{min} = 22.2$ bits, showing little information deficiency and indicating that a single cognate site is sufficient to address a TF to a specific location in prokaryotes. However, the average information content of a multicellular eukaryotic motif, 12.1 bits, falls far below $I_{min}$ ≈ 30 bits required to provide a specific address in a eukaryotic genome. Such significant information deficiency clearly demonstrates that a binding of a eukaryotic TF to a cognate site on DNA cannot be specific, i.e. many more equally strong sites are expected to be found and bound across the genome.

Information theory allows to estimate the expected number of such spurious motif matches, or hits, per genome. Using powerful Stein's lemma and simulations (Fig.S2) we demonstrate that the number of hits $h$ can be approximated and bound from below $h \geq 2^{I_{\min} - I}$ (Supplementary Methods). The average spacing $s$ between the hits is then approximated by $s \leq 2^I$ (Figure 1 and

Supplementary Methods). In a eukaryotic genome, $h \approx 10^4\text{-}10^6$, and even substantial chromatization (90%) only reduces $h$ to $10^3\text{-}10^5$ spurious binding sites per genome. Note that information content of a motif does not determine or constrain the number of cognate, functional sites a TF has in the genome. However, the large number of spurious high-affinity binding sites in genomes of multicellular eukaryotes creates a binding landscape with a potential for widespread non-functional binding. These information-theoretic estimates provide a lower bound on the number of spurious binding events. To test these estimates, we searched for matches to a several well-characterized motifs in real genomic sequences, and consistent with the theory, found many spurious hits to eukaryotic binding motifs (Supplementary Methods, Supplementary Table 1).

Our analysis shows that a single cognate site is sufficient to address a TF to a specific location in prokaryotes. In contrast, a typical eukaryotic TF is expected to have specific sites arising by chance every $s \approx 4000$ bps. The phenomenon of widespread non-functional binding has been recently observed experimentally for several TFs in *D. melanogaster* (Li *et al*, 2008) and *S. cerevisiae* (Hu *et al*). In fact, our estimate of $\sim 10^3$ spurious hits in the chromatinized fly genome agrees well with experimentally observed $10^3\text{-}10^4$ binding events. Moreover, our results help to explain the large number of binding events detected by ChIP-chip (Harbison *et al*), suggesting that the vast majority of these event reflect the widespread *specific* (high-affinity) binding of eukaryotic TFs to sites that are not functional, but inevitable appear in the genomic background. . In agreement with our results, a recent study of yeast TFs demonstrated little overlap between genes that are differentially expressed in response to TF knockout and targets bound by this TF, suggesting there are an abundance of binding events with no detectable effect on gene expression (Hu *et al*). The prevalence of widespread, unavoidable, spurious binding events in eukaryotes calls for caution in interpreting all experimentally identified binding events as regulatory interactions.

The abundance of accessible high-affinity spurious sites has two effects: (i) it sequesters TF molecules, and (ii) it makes harder for the cellular machinery of gene regulation (RNA polymerase, general transcription factors, etc.) to discriminate between functionally and non-functionally bound TFs. How then is addressing achieved in eukaryotes? We suggest that clustering of binding sites in regulatory regions combined with a sufficiently high copy-number of TFs allows eukaryotes to cope with the low information content of their TF motifs.

The sequestration of TF molecules by spurious binding sites necessitates many more TF copies per cell to occupy a few cognate sites. The number of spurious sites $h$ imposes a lower limit on the TF copy number per cell, which is approximately 5 copies per cell for bacteria, 2000 for yeast, and $10^3\text{-}10^5$ for multicellular eukaryotes (taking into account 90% chromatization). These estimates are remarkably consistent with available experimental data: 5-10 copies per cell of *Lac* repressor in *E. coli*, an average of approximately 2000 copies per cell of TFs in yeast and $10^5$ copies per cell of such prototypical multicellular eukaryotic TF as p53 (see Supplementary Table 4). Although high TF copy numbers are necessary to cope with non-functional binding, they are not sufficient to provide specificity, i.e. to allow cellular machinery to distinguish functional regulatory binding events from equally strong decoys.

Many regulatory regions in eukaryotes are known to contain multiple sites of the same or different TFs (e.g. (Ochoa-Espinosa *et al*, 2005)), suggesting that the clustering of sites can be used to predict such regions (Berman *et al*, 2004; Emberly *et al*, 2003; Siggia, 2005). Our analysis allows us to calculate the minimal number of immediately adjacent sites (*n*) needed to specify a unique location in a genome as $n = I_{min}/I \approx 3$ for multicellular eukaryotes and $n \approx 1$ for prokaryotic motifs. If sites are not immediately adjacent, but are located within a regulatory region of 500-1000 bps, more sites are needed to make a regulatory region stand out from the spurious binding event background. Using estimated the background frequency of hits for a single TF, we calculate the minimal number of sites of this TF per cluster ($n_{cluster}$) as 7-9 for a eukaryotic regulatory region of about 1000 bp (Supplementary Methods). Such a cluster of sites is expected to appear less than once per genome due to the spurious hits, thus allowing a cell to uniquely identify a regulatory region. If a regulatory region is composed of the sites of several different TFs, then the cluster should contain at least 12-20 binding sites in a regulatory region of 1000 bps (Supplementary Methods). This lower bound on the number of required binding sites is remarkably consistent with 20-25 sites per kilobase observed in fly developmental enhancers(Berman *et al*, 2004). While fly and sea urchin enhancers are known to contain clusters of TF binding sites, our results clearly demonstrate that clustering of sites should be a common phenomenon applicable to most regulatory regions and promoters in multicellular eukaryotes. Simply put, since a single eukaryotic binding motif is unable to specify a unique address in the genome, multiple binding sites must be used in order to unambiguously specify a regulation site.

What are the advantages that low-information TF motifs provide to a eukaryotic cell? First, the required clustering of sites provides a mechanism for combinatorial gene regulation: to be recognizable by a cell, a regulatory region should contain several TFs bound in close proximity, providing an opportunity for synergistic action between TFs. Second, the short motifs of eukaryotic TFs facilitate the rapid creation of new sites and rearrangement of existing sites(Berg *et al*, 2004), thus enabling highly evolvable gene regulation (Carroll, 2005) and the rapid turnover of sites in regulatory regions(Berman *et al*, 2004; Moses *et al*, 2006). And third, combinatorial regulation obtained through site clustering allows a large number of genes to be controlled by a limited repertoire of TFs (Itzkovitz *et al*, 2006). Our study shows that combinatorial regulation is rooted in the way eukaryotic TFs recognize DNA.

The observed difference in genome addressing strategy may have arisen in several different ways: gradual modifications of the DNA-binding residues of TFs, the expansion or contraction of the DNA-binding interface of TFs, the preferential use of one type of DNA-binding domain (e.g. zinc fingers) over others on a kingdom-wide scale, or the re-invention of DNA-binding domains altogether. To investigate the evolutionary trajectory of eukaryotic DNA recognition, we systematically compared sequences of prokaryotic and eukaryotic DNA-binding domains of TFs available in the PFAM database (Figure 2A).

This analysis gave a surprising result – prokaryotes and eukaryotes use different sets of DNA-binding domains. Of the 133 known DNA-binding domain families, 69 have only eukaryotic members, 49 are totally prokaryotic, and only 15 families with both prokaryotic and eukaryotic members, but are usually dominated by one of two kingdoms (Supplementary Methods, Supplementary Table 2). As a positive control, we compared this result to the domains involved in glycolysis and gluconeogenesis and found that a very small number of those domains are kingdom specific (Figure 2B). The lack of shared prokaryotic and eukaryotic DNA-binding domain families suggests that the TF machinery of low-specificity binding and largely combinatorial regulation employed by eukaryotes has evolved *de novo*.

This evolutionary analysis supports our information-theoretical results and emphasizes that the observed differences in DNA recognition are not specific to a few well-characterized TFs or organisms, but are likely to span across kingdoms and constitute fundamentally different strategies to transcriptional regulation in prokaryotes and eukaryotes.

## Materials and Methods

Binding motifs for *Escherichia coli* were downloaded from RegulonDB, yeast transcription factor motifs were taken from MacIsaac, et al. (MacIsaac *et al*, 2006), and the JASPAR CORE collection of eukaryotic transcription factor binding motifs was downloaded from JASPAR. Background nucleotide frequencies were calculated for those organisms with completed genome sequences and the average background frequencies were used for all others. Tables with the information content and GC content for all the transcription factors are in the Supplementary Data. The differences between prokaryotic and eukaryotic binding motifs are evident when using other motif collections (Supplementary Table 3).

The number of spurious binding events is estimated using both information-theoretical bound provided by Stein's Lemma and by simulations where the frequency of a motif with a given length and information content was calculated in a random DNA sequence of given background composition. Details are provided in the Supplementary Methods.

In our analysis of the composition of PFAM DNA-binding domain families, we eliminated the weakest 10% of matches to the PFAM domain model, which we call "filtering" in Supplementary Table 2. Without filtering, there are 67 eukaryotic domain families, 41 prokaryotic domain families and 25 families with both prokaryotic and eukaryotic members.

## Acknowledgments

We thank Shamil Sunyaev, Mikahil Gelfand, Shaun Mahoney and Alex Shpunt for insightful discussions and Michael Schnall for interpretation of the information cutoff. ZW was supported by a Howard Hughes Medical Institute Predoctoral Fellowship. LM acknowledges support of *i2b2*, NIH-supported Center for Biomedical Computing at the Brigham and Women's Hospital.

# Supplementary Materials

## Derivation of Hit Number Estimates

*Problem 1:* Given the information content $I$ of a motif and the length of the genome $N$, what is the expected number of sequences that can be classified as an instance of this motif (i.e. the number of hits)?

To solve this problem we introduces a quantity $P_{FM}$, *the probability of a false motif.* Given $P_{FM}$, the expectation of the number of hits per genome is simply

$$h = NP_{FM} \approx P_{FM} 2^{-I_{min}}.$$

*Problem 2:* Give a motif, i.e. frequencies of base-pairs at individual positions: $p_i(b) : i = 1...L, \ b \in \{A,C,G,T\}$, estimate the probability $P_{FM}$ of finding a *false motif* among sequences generated by the background frequency $q(b)$, i.e. the probability of finding a random genomic sequence that is classified as a motif.

The later problem is similar to a classical hypothesis testing (detection) problem: given two distributions $P(x)$ and $Q(x)$, classify an *i.i.d.* sample $\{x_1, x_2...x_L\}$ of as being an instance from either of the distributions (**Fig S1A**). Fundamental results of hypothesis testing theory can be immediately applied to motif recognition. Detection problem has two types of errors: a false alarm (type I) and a miss (type II). These correspond to two types of errors in site recognition: a *false motif*, i.e. a random sequence classified as a motif; and a *miss*, an instance of a motif classified as a random sequence. A common formulation of the detection problem is to constrain one type of an error and to minimize the other. Here we set $\Pr\{miss\} = \alpha$ and seek to estimate $P_{FM}$. No closed form solution for $P_{FM}$ is expected, but a theory can provide useful bounds that we test by simulations.

First we consider a few illustrative examples.

### Example 1: consensus sequence

If a transcription factor bound only one consensus sequence, $L$ nucleotides long, the information content of the binding motif is:

$$I = \sum_{i=1}^{L} \sum_{b \in \{A,C,G,T\}} p_i(b) \log_2 \frac{p_i(b)}{q(b)} = \sum_{i=1}^{L} 1 \cdot \log_2 \frac{1}{q(a(i))} = -\sum_{i=1}^{L} \log_2 q(a(i)),$$

where $a(i)$ is the nucleotide in position $i$ of the consensus (**Fig S1B**). The probability of a *false site* is then simply the probability of the consensus sequence to be found in the background:

$$P_{FM} = \Pr\{\text{consensus}\} = \prod_{i=1}^{L} q(a(i)) = 2^{-I}$$

Notice that the probability of a miss $\Pr\{miss\} = \alpha = 0$ in this case.

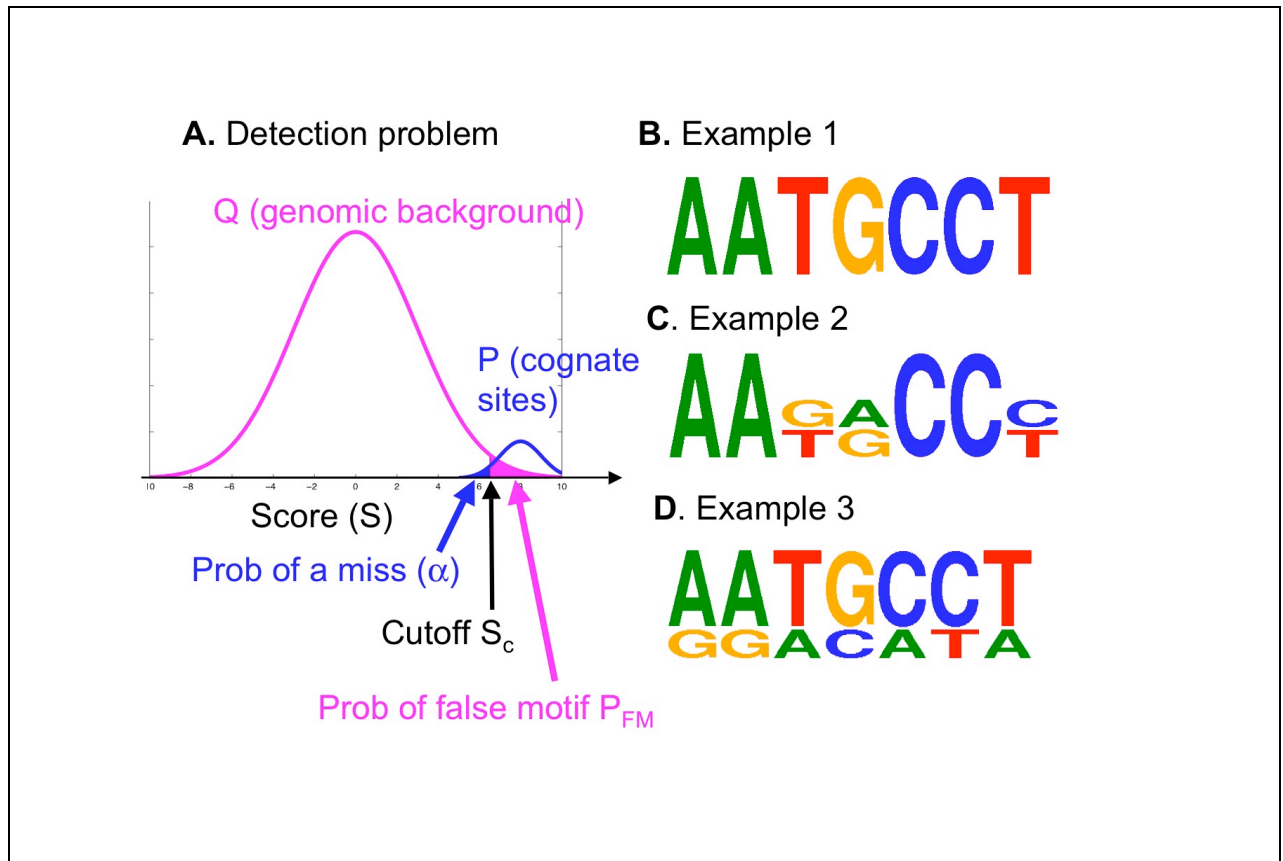

**Figure S1. A.** Illustration of the detection problem and two types of errors. **B-D.** Motifs from examples 1-3 in the Supplementary text.

*Example 2: consensus with 2-fold degenerate positions*
Consider a motif that has $l_1$ consensus positions and $l_2 = L - l_1$ two-fold degenerate positions, i.e. one of the two possible base-pairs in these positions. A site is said to be an instance of this motif if it has consensus base-pairs in the consensus positions and any of the two allowed base-pairs in the two-fold degenerate ones **(see Fig S1C)**: $p_i(b^{(1)}) = p_i(b^{(2)}) = 1/2$ for allowed base-pairs $b^{(1)}$ and $b^{(1)}$ in the two-fold degenerate position. To simplify notation we assume that $q(A) = q(C) = q(G) = q(T) = q$. In this case the information content is

$$I = \sum_{i=1}^{L} \sum_{b \in \{A,C,G,T\}} p_i(b) \log_2 \frac{p_i(b)}{q} = -l_1 \log_2 q - l_2 \log_2 2q,$$

while the probability of finding such site among background sequences is given by

$$P_{FM} = q^{l_1} (2q)^{l_2} = 2^{-I}.$$

If the background probabilities $q(a)$ are all different, we obtain the following:

$$I = -\sum_{i \in \{\text{consensus}\}}^{l_1} \log_2 q(a(i)) - \frac{1}{2} \sum_{i \in \{\text{degenarate}\}}^{l_2} \left( \log_2 2q(a^{(1)}(i)) + \log_2 2q(a^{(2)}(i)) \right)$$

$$P_{FM} = \prod_{i \in \{cons\}}^{l_1} \log_2 q(a(i)) \prod_{i \in \{degen\}}^{l_2} \left[ q(a^{(1)}(i)) + q(a^{(2)}(i)) \right] \geq \prod_{i \in \{cons\}}^{l_1} \log_2 q(a(i)) \prod_{i \in \{degen\}}^{l_2} 2\sqrt{q(a^{(1)}(i))q(a^{(2)}(i))} = 2^{-I}$$

The inequality follows from the famous inequality for the means: $(a+b)/2 \geq \sqrt{ab}$. Thus we obtain $P_{FM} \geq 2^{-I}$, i.e. $2^{-I}$ provides a lower bound for $P_{FM}$.

*Example 3: consensus with sub-optimal base-pairs*
Now consider a motif where each position is two-fold degenerate with the frequencies $p$ and $1-p$ of the two base-pairs appearing in cognate sites. Naturally, the identities of allowed base-pairs can be different at different positions of the site (**Fig S1D**). Again, to simplify notation we assume all background frequencies $q$ to be equal. The information content and the probability of a false motif are as following:

$$I = \sum_{i=1}^{L} \sum_{b \in \{A,C,G,T\}} p_i(b) \log_2 \frac{p_i(b)}{q} = L\left( p \log_2 \frac{p}{q} + (1-p) \log_2 \frac{1-p}{q} \right) = -LH(p) - L \log q,$$

where $H(p) = -p \log_2 p - (1-p) \log_2 (1-p)$ is the entropy of $p$.

$$P_{FM} = (2q)^L$$

Using convexity of the entropy $H(p) \leq H(1/2) = \log 2$, i.e. $\log 2 - H(p) > 0$, we obtain

$$P_{FM} = (2q)^L = 2^{-I+L[\log 2 - H(p)]} \geq 2^{-I},$$

with the equality at $p = 1/2$. Thus $2^{-I}$ provides a lower bound of the false motif probability. Thus the actual number of false motifs found shall be greater than $N 2^{-I}$.

*Stein's Lemma*
In the general case, the probability $P_{FM}$ is the second type error of the detection problem. An important *Stein's Lemma* provides the asymptotic value of this type of error, when the first type error (the miss probability) $\alpha \to 0$. For simplicity, consider a motif where $p_{i=1...L}(b) = p(b)$ are the same for all positions but can be different for different base-pairs, and the background frequencies $q(b)$ can be different as well. Then the information content of the motif is

$$I = L \sum_{b \in \{A,C,G,T\}} p(b) \log_2 \frac{p(b)}{q} \equiv LD(p \| q),$$

where the sum $D(p\|q)$ is the Kullback-Leibler divergence between $q$ and $p$. Stein's Lemma states that

$$P_{FM} \to 2^{-LD(p\|q)} = 2^{-I} \text{ for } \alpha \to 0, L \to \infty.$$

Moreover Cover and Thomas showed that $P_{FM} \geq 2^{-LD(p\|q)}(1-\alpha)$. It is straight forward to show that both the inequality and the limit hold for any form of the motif $p_{i=1...L}(b)$.

<u>In summary</u>, a few illustrative examples and Stein's Lemma show that the probability of a false motif instance is tightly bound below by $2^{-I}$. The quality of this bound depends on the length of the motif $L$, the acceptable miss probability $\alpha$, and other specifics of the problem. We test this bound by *simulations* and by the *bioinformatic analysis* presented below.

## Simulations

The goal of the simulations is to estimate the probability $P_{FM}$ of finding a false site among random background sequence, and to compare obtained $P_{FM}$ with its lower bound $2^{-I}$ calculated using motif's information content $I$.

We set background frequencies $q(b)$, and generate a motif $p_{i=1...L}(b)$ such that is has information content $I$ in a desirable range. Next we simulate a bioinformatic study that uses this motif to discover cognate sites. We generate a random "genomic" sequences using the background model $q(b)$ and calculate the probability of a false site in this genome. All discovered sites are false since no cognate sites have been planted into this random genome.

Specific steps of the algorithm are as follows:
1. Set $q(b)$ and generate $p_{i=1...L}(b)$ to obtain sought $I$.
2. Calculate the PWM $S_i(b) = \log p_i(b)/q(b)$
3. Set acceptable miss probability $\alpha\,(=0.05-0.15)$ and determine the cutoff $S_c$ that provides this miss probability. To find $S_c$, generate random cognate sites according to $p_{i=1...L}(b)$, calculate the score using the PWM, and find the cutoff $S_c$ such that $\alpha$ fraction of generated cognate sites have $S > S_c$ (see **Fig S1A**).
4. Generate 10000 background genomic sequences according to $q(b)$, calculate the score $S$ for each, and estimate the probability of a false motif as the fraction of random sequences that have $S > S_c$. Compare obtained $P_{FM}$ with its estimate and bound $2^{-I}$.

Figure S2 presents results of these simulations for different L, $\alpha$, and motif $p_{i=1...L}(b)$ that have different information contents. First, we see that $2^{-I}$ provides a good estimate and lower bound for $P_{FM}$ for a broad range of parameters. Second we see that $P_{FM}$ exceeds $2^{-I}$ by a factor of $2...8$ for most of motif length and information contents.

These simulations validate the use of $2^{-I}$ as the lower bound of the frequency of false site, thus supporting analysis presented in the paper. The value of $P_{FM}$ in excess of $2^{-I}$ suggests that discussed widespread non-functional binding may have even greater scope than estimated in the paper.

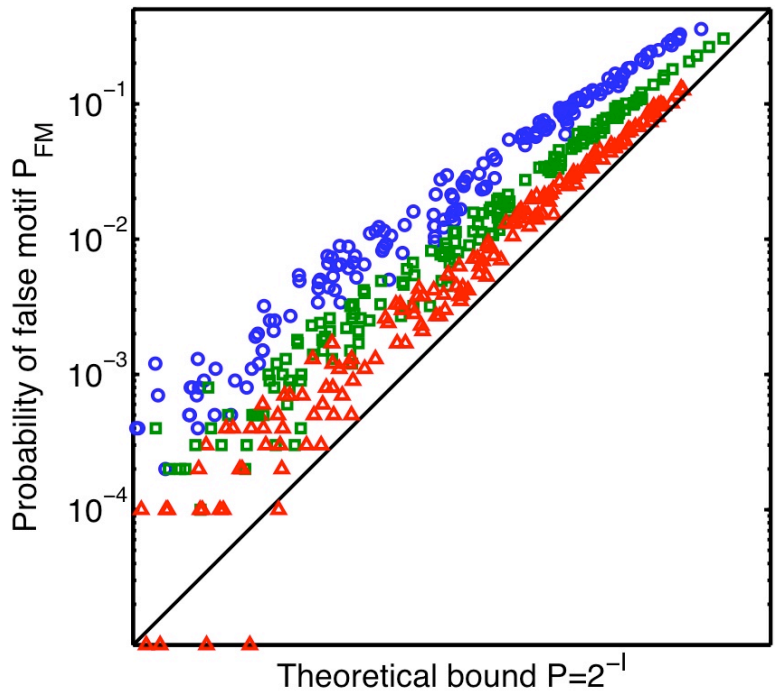

**Figure S2.** Simulations of motif recognition in random DNA sequences using PWM generated to have desired information content. Simulations demonstrate good agreement between the probability of a false motif and its theoretical lower bound. Generated motifs have lengths L=8,12,20 and the information content in a range of 0.6-1.2 bits per base pair.

## *Bioinformatics*

To check the validity of our estimates for the expected number of hits to a given motif in a genome, we looked at several examples. We selected five TF binding motifs, and calculated the information content of each. We have chosen TFs for which functional cognate sites are well-known, thus allowing us to discriminate between functional (known) and mostly non-functional sites discovered in the genome.

We created a position weight matrix for each using the log ratio of nucleotide frequencies in the motif and the genomic background. Then, for each TF, we downloaded a list of "real" binding sites. For yeast TFs, the sites are from the *Saccharomyces* Genome Database (http://www.yeastgenome.org/), for *Drosophlia* TFs, from D. Papatsenko's FlyDev resource website (http://flydev.berkeley.edu/cgi-bin/Annotation/), and for *E. coli*, from RegulonDB[2]. For each set, we aligned the sites to PWM, adding and deleting consensus letters from the sites to ensure the PWM and sites matched. We chose a score cutoff that qualified a stretch of nucleotides as a "hit" based on the scores of the "real" sites – ensuring that real sites would be considered hits. We then used the PATSER program (http://ural.wustl.edu/software.html) to calculate the number of hits to each PWM against the whole yeast, fly or *E. coli* genome.

The results of this analysis are presented in **Table S1**. As expected from the simulations, the number of observed hits exceeds $N2^{-I}$ estimate by a factor of 4-6.

## Supplementary Table 1.
## Number of expected and actual hits to TF PWMs

| TF | Organism | Information Content $I$ | Genome Size $N$ | Expected Hits $h = N2^{-I}$ | Actual Hits | Known cognate sites |
|---|---|---|---|---|---|---|
| CBF1 | Yeast | 13.97 | 1.20E+07 | 1496 | 9759 | 123 |
| GCN4 | Yeast | 10.66 | 1.20E+07 | 14833 | 115689 | 130 |
| Dorsal_1 | Fly | 13.39 | 1.37E+08 | 25525 | 150301 | ? |
| LacI | E. coli | 31.18 | 5.00E+06 | 0 | 3 | 3 |
| GalS | E. coli | 20.38 | 5.00E+06 | 7 | 121 | 7 |

## *Estimate of Yeast Transcription Factor Copy Numbers*

To estimate the average copy number of transcription factors in yeast, we used two different datasets:
1. We used Saccharomyces Genome Database (SGD) to obtain a list of genes annotated by Genome Ontology term "transcription factor activity" (GO:0003700). Both computationally predicted (118) and manually curated (63 genes) lists were used. The copy number data were obtained from YeastGFP database, resulting in 117 TFs with an assigned YeastGFP signal and 97 having a numerical estimate of the copy number. The mean and the median are 1567 and 704 copies/cell, respectively.
2. The data set published by Lu, et al[3] was used. We extracted the copy number estimates for the 33 yeast transcription factors included in both our data set and Lu, et al., and found the average copy number to be 2238 copies/cell.

## *Calculations for $n_{cluster}$*

Here we calculate the minimal number of binding sites per cluster needed to make it stand out from the background of spurious binding sites. We assume cluster to be localized within a region of $w$ (300-1000bps). The idea is to find the minimal number $n_{cluster}$ of spurious site in a cluster, such that the expected number of such cluster in the whole genome is less than 1. Given the probability of a spurious hit $p = P_{FM}$ (see above and Table 1) and the window size, the Poisson probability of observing $k$ hits in a window is:

$$P(k) = \binom{w}{k} p^k (1-p)^{w-k} \approx \frac{\lambda^k}{k!} e^{-\lambda},$$

where $\lambda = pw$, and the expected number of a window with $k$ hits in the genome of length $N$ is $E(k) = P(k) \cdot N$. We seek $n_{cluster}$ as the minimal value of $k$ for which $E(k) < 1$, or approximately $\log E(n_{cluster}) \approx 0$.

A cluster of more than $n_{cluster}$ sites is unlikely to appear in a genome due to spurious sites, thus providing a lower bound on the number of sites in a functional (distinct) cluster. Using the following parameters we obtain estimates for $n_{cluster}$:

|  | Yeast | Yeast | Multicellular Eukaryotes |
|---|---|---|---|
| $w$ | 500 bps | 1000 bps | 1000 bps |
| $p$ | $1/(6 \cdot 10^3)$ | $1/(6 \cdot 10^3)$ | $1/(4 \cdot 10^3)$ |
| $N$ | $1.2 \cdot 10^7$ bps | $1.2 \cdot 10^7$ bps | $10^8$-$10^{10}$ bps |
| **$n_{cluster}$** | 5 | 6 | 7-9 |

If a cluster is composed of sites of several different TFs, then the minimal number of sites to form a distinct cluster is different. Let $m$ be the number of TFs whose sites form a cluster. If the probability of a spurious site of an individual TF is $p$, then the probability of a spurious site of *any* of these TFs is approximately $p \cdot m$. Using $\lambda = pmw$ in equation for $P(k)$ we obtain:

|   | Yeast | Yeast | Multicellular Eukaryotes | Multicellular Eukaryotes |
|---|---|---|---|---|
| $m$ | 2 | 4 | 3 | 10 |
| $w$ | 500 bps | 1000 bps | 1000 bps | 1000 bps |
| $p$ | $1/(6 \cdot 10^3)$ | $1/(6 \cdot 10^3)$ | $1/(4 \cdot 10^3)$ | $1/(4 \cdot 10^3)$ |
| $N$ | $1.2 \cdot 10^7$ bps | $1.2 \cdot 10^7$ bps | $10^8$-$10^{10}$ bps | $10^8$-$10^{10}$ bps |
| $n_{cluster}$ | **6** | **9** | **10-12** | **16-19** |

**Supplementary Table 1. Number of expected and actual hits to TF PWMs**

| TF | Organism | Information Content | Genome Size | Expected Hits | Acutal Hits | "Real" Hits in Genome |
|---|---|---|---|---|---|---|
| CBF1 | Yeast | 13.97 | 1.20E+07 | 1496 | 9759 | 123 |
| GCN4 | Yeast | 10.66 | 1.20E+07 | 14833 | 115689 | 130 |
| Dorsal_1 | Fly | 13.39 | 1.37E+08 | 25525 | 150301 | ? |
| LacI | E. coli | 31.18 | 5.00E+06 | 0 | 3 | 3 |
| GalS | E. coli | 20.38 | 5.00E+06 | 7 | 121 | 7 |

**Supplementary Table 2. PFAM DNA-binding domain families with hits to prokaryotes and eukaryotes**

| PFAM ID | Name | Excluded with filtering |
|---------|------|------------------------|
| PF00126 | Bacterial regulatory helix-turn-helix protein, lysR family | yes |
| PF00486 | Transcriptional regulatory protein, C terminal | no |
| PF04383 | KilA-N domain | no |
| PF01381 | Helix-turn-helix | no |
| PF02954 | Bacterial regulatory protein, Fis family | no |
| PF00313 | Cold-shock' DNA-binding domain | yes |
| PF00325 | Bacterial regulatory proteins, crp family | no |
| PF01047 | MarR family | no |
| PF04299 | Putative FMN-binding domain | yes |
| PF00392 | Bacterial regulatory proteins, gntR family | no |
| PF00165 | Bacterial regulatory helix-turn-helix proteins, AraC family | no |
| PF00096 | Zinc finger, C2H2 type | no |
| PF05225 | helix-turn-helix, Psq domain | yes |
| PF00847 | AP2 domain | no |
| PF04967 | HTH DNA binding domain | no |
| PF08279 | HTH domain | yes |
| PF01022 | Bacterial regulatory protein, arsR family | no |
| PF00196 | Bacterial regulatory proteins, luxR family | yes |
| PF00010 | Helix-loop-helix DNA-binding domain | no |
| PF00356 | Bacterial regulatory proteins, lacI family | yes |
| PF02082 | Transcriptional regulator | no |
| PF00292 | Paired box' domain | yes |
| PF04397 | LytTr DNA-binding domain | yes |
| PF03749 | Sugar fermentation stimulation protein | no |
| PF04353 | Regulator of RNA polymerase sigma70 subunit, Rsd/AlgQ | yes |

**Supplementary Table 3. Average information content from other data sources**

| Data Set | Organism | Mean Information Content | Number of Motifs |
|---|---|---|---|
| DP Interact | E. coli | 24.3 | 68 |
| Harbison, et al. | Yeast | 14.1 | 102 |
| FlyReg | Fly | 12.5 | 75 |
| Bergman, et al. | Fly | 13.4 | 62 |
| TRANSFAC 11.3 | Assorted | 13.0 | 834 |

These data sets were accessed from STAMP (http://www.benoslab.pitt.edu/stamp)
S Mahony, PV Benos, "STAMP: a web tool for exploring DNA-binding motif similarities", Nucleic Acids Research (2007) 35:W253-W258.

**Supplementary Table 4.** The number of TF copies per cell

| Transcription Factor | Organism | TF copies per cell | Source |
|---|---|---|---|
| *LacI* tetramer | *E.coli* | 10 | [1] |
| *LacI* dimers | *E.coli* | 3 | [2] |
| 119 TF as annotated in GO (lists obtained from SGD, CopyNumbers from YeastGFP) | *S.cerevisiae* | Mean 1600 Median 700 | [3] |
| p53 | *H.Sapiens* | $10^5$ | www.bioNumbers.org and [4] |

# Figure legends

*Figure 1.*  *Properties of binding sites for bacteria, yeast, and multicellular eukaryotes*
The bar chart displays the minimum required information content for bacteria, yeast, and multicellular eukaryotes (red), as well as the actual information content of TF binding motifs (blue) for 72 bacterial, 124 yeast and 123 multicellular eukaryotic motifs.  The error bars are ± 1 standard deviation for the actual information content, and for eukaryotic $I_{min}$, the error bars represent the variability in that quantity due to the range of genome sizes $N$.  Below each series in the bar chart, we display an example of sequence logo for a binding motif with close to average information content, and other important properties of TF binding motifs.

*Figure 2.*  *Membership of PFAM protein domain families, by kingdom*
To explore the evolution of TF DNA-binding domains, we examined the membership of PFAM protein domain families.  Each column of (a-b) represents a single PFAM family, and the size of the orange or teal bar indicates the fraction of the family's bacterial and eukaryotic members, respectively.   In (a), we plot the membership of DNA-binding domains (from DBD database(Wilson *et al*, 2007)), demonstrating that they are almost unshared by bacteria and eukaryotes, and in (c), we show a Venn diagram, after removing the weakest 10% of hits to a PFAM family profile.    As a control (b), we plot the composition of PFAM glycolysis/gluconeogensis enzyme families (as reported in KEGG database), which are shared by bacteria and eukaryotes.

**Figures**

Fig. 1

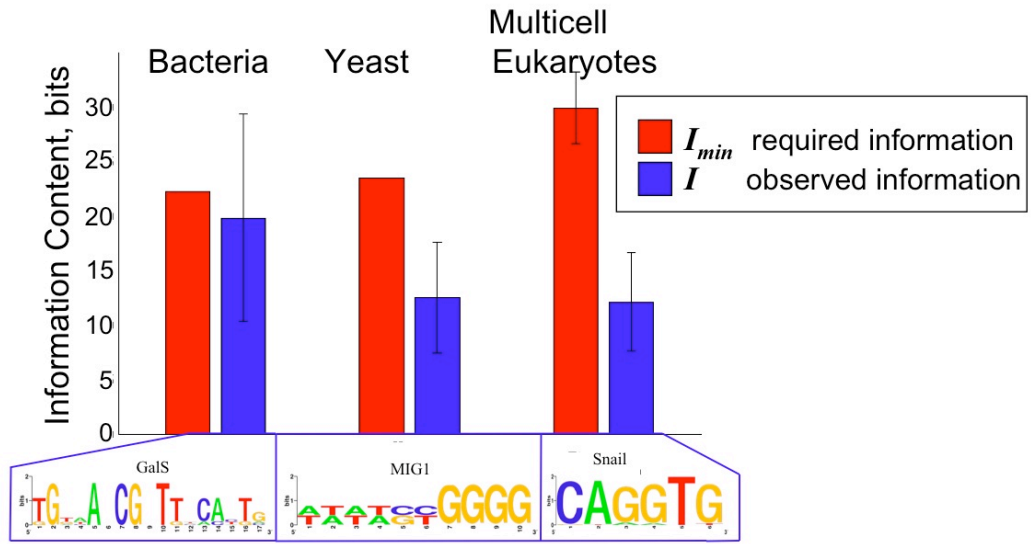

| | Bacteria | Yeast | Multicell Eukaryotes |
|---|---|---|---|
| Genome size ($N$) | $5 \cdot 10^6$ bps | $1.2 \cdot 10^7$ bps | $10^8 - 10^{10}$ bps |
| Expected number of spurious hits ($h$) | 5 | 2,000 | 20,000 – 2,000,000 |
| Spacing between hits ($s$) | $9 \times 10^5$ bps | 6,000 bps | 4,000 bps |
| Minimal number of adjacent sites ($n$) | 1 | 2 | 2-3 |
| Minimal number of sites per cluster of 1Kbs ($n_{cluster}$) | 1 | 4-6 | 7-9 |

Figure 2

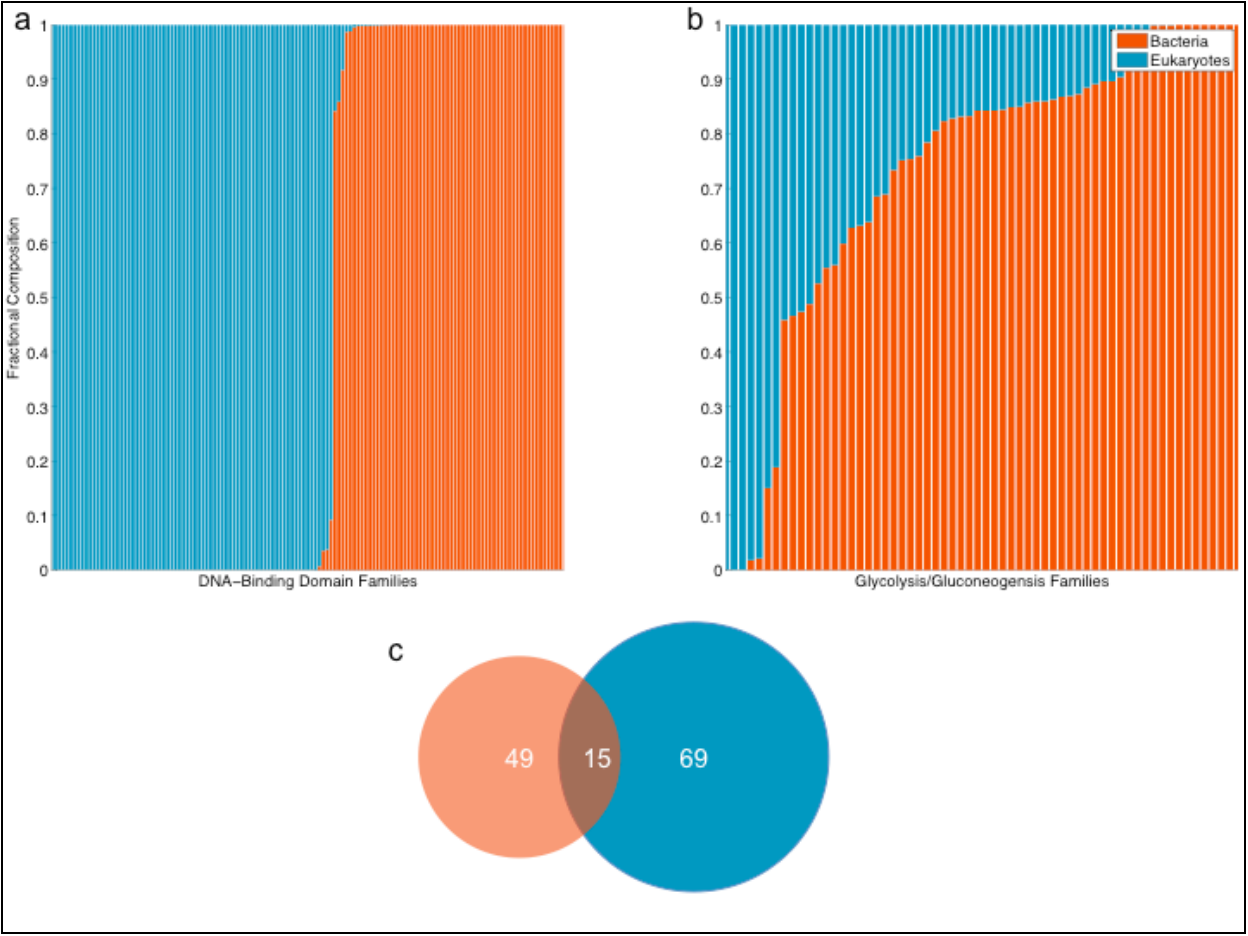

# Supplementary Materials

## Derivation of Hit Number Estimates

*Problem 1:* Given the information content $I$ of a motif and the length of the genome $N$, what is the expected number of sequences that can be classified as an instance of this motif (i.e. the number of hits)?

To solve this problem we introduces a quantity $P_{FM}$, *the probability of a false motif*. Given $P_{FM}$, the expectation of the number of hits per genome is simply

$$h = NP_{FM} \approx P_{FM} 2^{-I_{min}}.$$

*Problem 2:* Give a motif, i.e. frequencies of base-pairs at individual positions: $p_i(b): i = 1...L, \; b \in \{A,C,G,T\}$, estimate the probability $P_{FM}$ of finding a *false motif* among sequences generated by the background frequency $q(b)$, i.e. the probability of finding a random genomic sequence that is classified as a motif.

The later problem is similar to a classical hypothesis testing (detection) problem: given two distributions $P(x)$ and $Q(x)$, classify an *i.i.d.* sample $\{x_1, x_2...x_L\}$ of as being an instance from either of the distributions (**Fig S1A**). Fundamental results of hypothesis testing theory can be immediately applied to motif recognition. Detection problem has two types of errors: a false alarm (type I) and a miss (type II). These correspond to two types of errors in site recognition: a *false motif*, i.e. a random sequence classified as a motif; and a *miss*, an instance of a motif classified as a random sequence. A common formulation of the detection problem is to constrain one type of an error and to minimize the other. Here we set $\Pr\{miss\} = \alpha$ and seek to estimate $P_{FM}$. No closed form solution for $P_{FM}$ is expected, but a theory can provide useful bounds that we test by simulations.
First we consider a few illustrative examples.

### Example 1: consensus sequence
If a transcription factor bound only one consensus sequence, $L$ nucleotides long, the information content of the binding motif is:

$$I = \sum_{i=1}^{L} \sum_{b \in \{A,C,G,T\}} p_i(b) \log_2 \frac{p_i(b)}{q(b)} = \sum_{i=1}^{L} 1 \cdot \log_2 \frac{1}{q(a(i))} = -\sum_{i=1}^{L} \log_2 q(a(i)),$$

where $a(i)$ is the nucleotide in position $i$ of the consensus (**Fig S1B**). The probability of a *false site* is then simply the probability of the consensus sequence to be found in the background:

$$P_{FM} = \Pr\{consensus\} = \prod_{i=1}^{L} q(a(i)) = 2^{-I}$$

Notice that the probability of a miss $\Pr\{miss\} = \alpha = 0$ in this case.

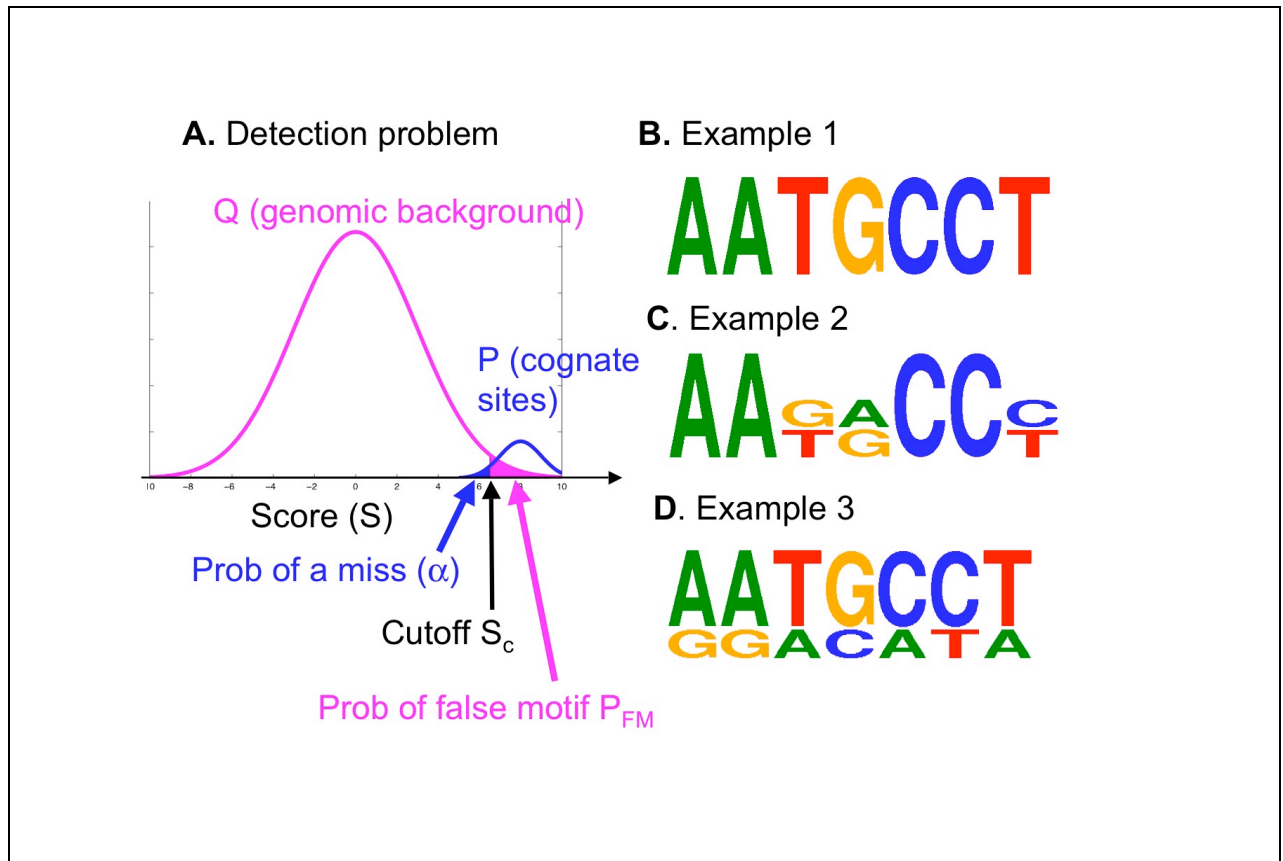

**Figure S1. A.** Illustration of the detection problem and two types of errors. **B-D.** Motifs from examples 1-3 in the Supplementary text.

*Example 2: consensus with 2-fold degenerate positions*
Consider a motif that has $l_1$ consensus positions and $l_2 = L - l_1$ two-fold degenerate positions, i.e. one of the two possible base-pairs in these positions. A site is said to be an instance of this motif if it has consensus base-pairs in the consensus positions and any of the two allowed base-pairs in the two-fold degenerate ones **(see Fig S1C)**: $p_i(b^{(1)}) = p_i(b^{(2)}) = 1/2$ for allowed base-pairs $b^{(1)}$ and $b^{(1)}$ in the two-fold degenerate position. To simplify notation we assume that $q(A) = q(C) = q(G) = q(T) = q$. In this case the information content is

$$I = \sum_{i=1}^{L} \sum_{b \in \{A,C,G,T\}} p_i(b) \log_2 \frac{p_i(b)}{q} = -l_1 \log_2 q - l_2 \log_2 2q,$$

while the probability of finding such site among background sequences is given by

$$P_{FM} = q^{l_1} (2q)^{l_2} = 2^{-I}.$$

If the background probabilities $q(a)$ are all different, we obtain the following:

$$I = -\sum_{i \in \{consensus\}}^{l_1} \log_2 q(a(i)) - \frac{1}{2} \sum_{i \in \{degenarate\}}^{l_2} \left( \log_2 2q(a^{(1)}(i)) + \log_2 2q(a^{(2)}(i)) \right)$$

$$P_{FM} = \prod_{i \in \{cons\}}^{l_1} \log_2 q(a(i)) \prod_{i \in \{degen\}}^{l_2} \left[ q(a^{(1)}(i)) + q(a^{(2)}(i)) \right] \geq \prod_{i \in \{cons\}}^{l_1} \log_2 q(a(i)) \prod_{i \in \{degen\}}^{l_2} 2\sqrt{q(a^{(1)}(i))q(a^{(2)}(i))} = 2^{-I}$$

The inequality follows from the famous inequality for the means: $(a+b)/2 \geq \sqrt{ab}$. Thus we obtain $P_{FM} \geq 2^{-I}$, i.e. $2^{-I}$ provides a lower bound for $P_{FM}$.

*Example 3: consensus with sub-optimal base-pairs*
Now consider a motif where each position is two-fold degenerate with the frequencies $p$ and $1-p$ of the two base-pairs appearing in cognate sites. Naturally, the identities of allowed base-pairs can be different at different positions of the site (**Fig S1D**). Again, to simplify notation we assume all background frequencies $q$ to be equal. The information content and the probability of a false motif are as following:

$$I = \sum_{i=1}^{L} \sum_{b \in \{A,C,G,T\}} p_i(b) \log_2 \frac{p_i(b)}{q} = L \left( p \log_2 \frac{p}{q} + (1-p) \log_2 \frac{1-p}{q} \right) = -LH(p) - L\log q,$$

where $H(p) = -p\log_2 p - (1-p)\log_2(1-p)$ is the entropy of $p$.

$$P_{FM} = (2q)^L$$

Using convexity of the entropy $H(p) \leq H(1/2) = \log 2$, i.e. $\log 2 - H(p) > 0$, we obtain

$$P_{FM} = (2q)^L = 2^{-I+L[\log 2 - H(p)]} \geq 2^{-I},$$

with the equality at $p = 1/2$. Thus $2^{-I}$ provides a lower bound of the false motif probability. Thus the actual number of false motifs found shall be greater than $N 2^{-I}$.

*Stein's Lemma*
In the general case, the probability $P_{FM}$ is the second type error of the detection problem. An important *Stein's Lemma* provides the asymptotic value of this type of error, when the first type error (the miss probability) $\alpha \to 0$. For simplicity, consider a motif where $p_{i=1...L}(b) = p(b)$ are the same for all positions but can be different for different base-pairs, and the background frequencies $q(b)$ can be different as well. Then the information content of the motif is

$$I = L \sum_{b \in \{A,C,G,T\}} p(b) \log_2 \frac{p(b)}{q} \equiv LD(p \| q),$$

where the sum $D(p\|q)$ is the Kullback-Leibler divergence between $q$ and $p$. Stein's Lemma states that

$$P_{FM} \to 2^{-LD(p\|q)} = 2^{-I} \text{ for } \alpha \to 0, L \to \infty.$$

Moreover Cover and Thomas showed that $P_{FM} \geq 2^{-LD(p\|q)}(1-\alpha)$. It is straight forward to show that both the inequality and the limit hold for any form of the motif $p_{i=1...L}(b)$.

In summary, a few illustrative examples and Stein's Lemma show that the probability of a false motif instance is tightly bound below by $2^{-I}$. The quality of this bound depends on the length of the motif $L$, the acceptable miss probability $\alpha$, and other specifics of the problem. We test this bound by *simulations* and by the *bioinformatic analysis* presented below.

## *Simulations*

The goal of the simulations is to estimate the probability $P_{FM}$ of finding a false site among random background sequence, and to compare obtained $P_{FM}$ with its lower bound $2^{-I}$ calculated using motif's information content $I$.

We set background frequencies $q(b)$, and generate a motif $p_{i=1...L}(b)$ such that is has information content $I$ in a desirable range. Next we simulate a bioinformatic study that uses this motif to discover cognate sites. We generate a random "genomic" sequences using the background model $q(b)$ and calculate the probability of a false site in this genome. All discovered sites are false since no cognate sites have been planted into this random genome.

Specific steps of the algorithm are as follows:
1. Set $q(b)$ and generate $p_{i=1...L}(b)$ to obtain sought $I$.
2. Calculate the PWM $S_i(b) = \log p_i(b)/q(b)$
3. Set acceptable miss probability $\alpha (=0.05-0.15)$ and determine the cutoff $S_c$ that provides this miss probability. To find $S_c$, generate random cognate sites according to $p_{i=1...L}(b)$, calculate the score using the PWM, and find the cutoff $S_c$ such that $\alpha$ fraction of generated cognate sites have $S > S_c$ (see **Fig S1A**).
4. Generate 10000 background genomic sequences according to $q(b)$, calculate the score $S$ for each, and estimate the probability of a false motif as the fraction of random sequences that have $S > S_c$. Compare obtained $P_{FM}$ with its estimate and bound $2^{-I}$.

Figure S2 presents results of these simulations for different L, $\alpha$, and motif $p_{i=1...L}(b)$ that have different information contents. First, we see that $2^{-I}$ provides a good estimate and lower bound for $P_{FM}$ for a broad range of parameters. Second we see that $P_{FM}$ exceeds $2^{-I}$ by a factor of $2...8$ for most of motif length and information contents.

These simulations validate the use of $2^{-I}$ as the lower bound of the frequency of false site, thus supporting analysis presented in the paper. The value of $P_{FM}$ in excess of $2^{-I}$ suggests that discussed widespread non-functional binding may have even greater scope than estimated in the paper.

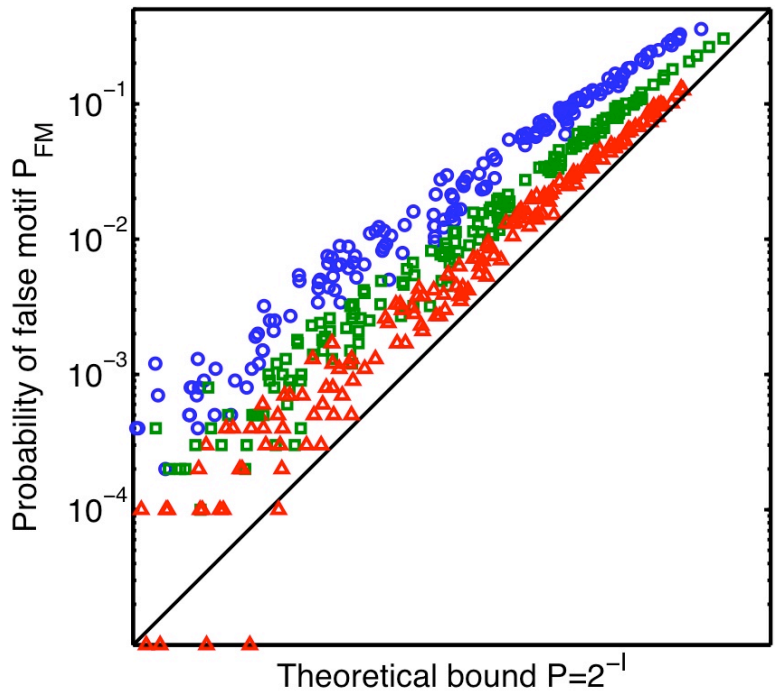

**Figure S2.** Simulations of motif recognition in random DNA sequences using PWM generated to have desired information content. Simulations demonstrate good agreement between the probability of a false motif and its theoretical lower bound. Generated motifs have lengths L=8,12,20 and the information content in a range of 0.6-1.2 bits per base pair.

## *Bioinformatics*

To check the validity of our estimates for the expected number of hits to a given motif in a genome, we looked at several examples. We selected five TF binding motifs, and calculated the information content of each. We have chosen TFs for which functional cognate sites are well-known, thus allowing us to discriminate between functional (known) and mostly non-functional sites discovered in the genome.

We created a position weight matrix for each using the log ratio of nucleotide frequencies in the motif and the genomic background. Then, for each TF, we downloaded a list of "real" binding sites. For yeast TFs, the sites are from the *Saccharomyces* Genome Database (http://www.yeastgenome.org/), for *Drosophlia* TFs, from D. Papatsenko's FlyDev resource website (http://flydev.berkeley.edu/cgi-bin/Annotation/), and for *E. coli*, from RegulonDB[2]. For each set, we aligned the sites to PWM, adding and deleting consensus letters from the sites to ensure the PWM and sites matched. We chose a score cutoff that qualified a stretch of nucleotides as a "hit" based on the scores of the "real" sites – ensuring that real sites would be considered hits. We then used the PATSER program (http://ural.wustl.edu/software.html) to calculate the number of hits to each PWM against the whole yeast, fly or *E. coli* genome.

The results of this analysis are presented in **Table S1**. As expected from the simulations, the number of observed hits exceeds $N2^{-I}$ estimate by a factor of 4-6.

## Supplementary Table 1.
## Number of expected and actual hits to TF PWMs

| TF | Organism | Information Content $I$ | Genome Size $N$ | Expected Hits $h = N2^{-I}$ | Actual Hits | Known cognate sites |
|---|---|---|---|---|---|---|
| CBF1 | Yeast | 13.97 | 1.20E+07 | 1496 | 9759 | 123 |
| GCN4 | Yeast | 10.66 | 1.20E+07 | 14833 | 115689 | 130 |
| Dorsal_1 | Fly | 13.39 | 1.37E+08 | 25525 | 150301 | ? |
| LacI | E. coli | 31.18 | 5.00E+06 | 0 | 3 | 3 |
| GalS | E. coli | 20.38 | 5.00E+06 | 7 | 121 | 7 |

## *Estimate of Yeast Transcription Factor Copy Numbers*

To estimate the average copy number of transcription factors in yeast, we used two different datasets:

1. We used Saccharomyces Genome Database (SGD) to obtain a list of genes annotated by Genome Ontology term "transcription factor activity" (GO:0003700). Both computationally predicted (118) and manually curated (63 genes) lists were used. The copy number data were obtained from YeastGFP database, resulting in 117 TFs with an assigned YeastGFP signal and 97 having a numerical estimate of the copy number. The mean and the median are 1567 and 704 copies/cell, respectively.
2. The data set published by Lu, et al[3] was used. We extracted the copy number estimates for the 33 yeast transcription factors included in both our data set and Lu, et al., and found the average copy number to be 2238 copies/cell.

## *Calculations for $n_{cluster}$*

Here we calculate the minimal number of binding sites per cluster needed to make it stand out from the background of spurious binding sites. We assume cluster to be localized within a region of $w$ (300-1000bps). The idea is to find the minimal number $n_{cluster}$ of spurious site in a cluster, such that the expected number of such cluster in the whole genome is less than 1. Given the probability of a spurious hit $p = P_{FM}$ (see above and Table 1) and the window size, the Poisson probability of observing $k$ hits in a window is:

$$P(k) = \binom{w}{k} p^k (1-p)^{w-k} \approx \frac{\lambda^k}{k!} e^{-\lambda},$$

where $\lambda = pw$, and the expected number of a window with $k$ hits in the genome of length $N$ is $E(k) = P(k) \cdot N$. We seek $n_{cluster}$ as the minimal value of $k$ for which $E(k) < 1$, or approximately $\log E(n_{cluster}) \approx 0$.

A cluster of more than $n_{cluster}$ sites is unlikely to appear in a genome due to spurious sites, thus providing a lower bound on the number of sites in a functional (distinct) cluster. Using the following parameters we obtain estimates for $n_{cluster}$:

|  | Yeast | Yeast | Multicellular Eukaryotes |
|---|---|---|---|
| $w$ | 500 bps | 1000 bps | 1000 bps |
| $p$ | $1/(6 \cdot 10^3)$ | $1/(6 \cdot 10^3)$ | $1/(4 \cdot 10^3)$ |
| $N$ | $1.2 \cdot 10^7$ bps | $1.2 \cdot 10^7$ bps | $10^8$-$10^{10}$ bps |
| **$n_{cluster}$** | 5 | 6 | 7-9 |

If a cluster is composed of sites of several different TFs, then the minimal number of sites to form a distinct cluster is different. Let $m$ be the number of TFs whose sites form a cluster. If the probability of a spurious site of an individual TF is $p$, then the probability of a spurious site of *any* of these TFs is approximately $p \cdot m$. Using $\lambda = pmw$ in equation for $P(k)$ we obtain:

|  | Yeast | Yeast | Multicellular Eukaryotes | Multicellular Eukaryotes |
|---|---|---|---|---|
| $m$ | 2 | 4 | 3 | 10 |
| $w$ | 500 bps | 1000 bps | 1000 bps | 1000 bps |
| $p$ | $1/(6 \cdot 10^3)$ | $1/(6 \cdot 10^3)$ | $1/(4 \cdot 10^3)$ | $1/(4 \cdot 10^3)$ |
| $N$ | $1.2 \cdot 10^7$ bps | $1.2 \cdot 10^7$ bps | $10^8$-$10^{10}$ bps | $10^8$-$10^{10}$ bps |
| $n_{cluster}$ | **6** | **9** | **10-12** | **16-19** |

**Supplementary Table 1. Number of expected and actual hits to TF PWMs**

| TF | Organism | Information Content | Genome Size | Expected Hits | Acutal Hits | "Real" Hits in Genome |
|---|---|---|---|---|---|---|
| CBF1 | Yeast | 13.97 | 1.20E+07 | 1496 | 9759 | 123 |
| GCN4 | Yeast | 10.66 | 1.20E+07 | 14833 | 115689 | 130 |
| Dorsal_1 | Fly | 13.39 | 1.37E+08 | 25525 | 150301 | ? |
| LacI | E. coli | 31.18 | 5.00E+06 | 0 | 3 | 3 |
| GalS | E. coli | 20.38 | 5.00E+06 | 7 | 121 | 7 |

**Supplementary Table 2. PFAM DNA-binding domain families with hits to prokaryotes and eukaryotes**

| PFAM ID | Name | Excluded with filtering |
|---|---|---|
| PF00126 | Bacterial regulatory helix-turn-helix protein, lysR family | yes |
| PF00486 | Transcriptional regulatory protein, C terminal | no |
| PF04383 | KilA-N domain | no |
| PF01381 | Helix-turn-helix | no |
| PF02954 | Bacterial regulatory protein, Fis family | no |
| PF00313 | Cold-shock' DNA-binding domain | yes |
| PF00325 | Bacterial regulatory proteins, crp family | no |
| PF01047 | MarR family | no |
| PF04299 | Putative FMN-binding domain | yes |
| PF00392 | Bacterial regulatory proteins, gntR family | no |
| PF00165 | Bacterial regulatory helix-turn-helix proteins, AraC family | no |
| PF00096 | Zinc finger, C2H2 type | no |
| PF05225 | helix-turn-helix, Psq domain | yes |
| PF00847 | AP2 domain | no |
| PF04967 | HTH DNA binding domain | no |
| PF08279 | HTH domain | yes |
| PF01022 | Bacterial regulatory protein, arsR family | no |
| PF00196 | Bacterial regulatory proteins, luxR family | yes |
| PF00010 | Helix-loop-helix DNA-binding domain | no |
| PF00356 | Bacterial regulatory proteins, lacI family | yes |
| PF02082 | Transcriptional regulator | no |
| PF00292 | Paired box' domain | yes |
| PF04397 | LytTr DNA-binding domain | yes |
| PF03749 | Sugar fermentation stimulation protein | no |
| PF04353 | Regulator of RNA polymerase sigma70 subunit, Rsd/AlgQ | yes |

**Supplementary Table 3. Average information content from other data sources**

| Data Set | Organism | Mean Information Content | Number of Motifs |
| --- | --- | --- | --- |
| DP Interact | E. coli | 24.3 | 68 |
| Harbison, et al. | Yeast | 14.1 | 102 |
| FlyReg | Fly | 12.5 | 75 |
| Bergman, et al. | Fly | 13.4 | 62 |
| TRANSFAC 11.3 | Assorted | 13.0 | 834 |

These data sets were accessed from STAMP (http://www.benoslab.pitt.edu/stamp)
S Mahony, PV Benos, "STAMP: a web tool for exploring DNA-binding motif similarities", Nucleic Acids Research (2007) 35:W253-W258.

**Supplementary Table 4.** The number of TF copies per cell

| Transcription Factor | Organism | TF copies per cell | Source |
|---|---|---|---|
| *LacI* tetramer | *E.coli* | 10 | [1] |
| *LacI* dimers | *E.coli* | 3 | [2] |
| 119 TF as annotated in GO (lists obtained from SGD, CopyNumbers from YeastGFP) | *S.cerevisiae* | Mean 1600 Median 700 | [3] |
| p53 | *H.Sapiens* | $10^5$ | www.bioNumbers.org and [4] |

1. Kao-Huang, Y. et al. Nonspecific DNA binding of genome-regulating proteins as a biological control mechanism: measurement of DNA-bound Escherichia coli lac repressor in vivo. *Proc Natl Acad Sci U S A* **74**, 4228-32 (1977).
2. Elf, J., Li, G.W. & Xie, X.S. Probing transcription factor dynamics at the single-molecule level in a living cell. *Science* **316**, 1191-4 (2007).
3. Chen, S.C., Zhao, T., Gordon, G.J. & Murphy, R.F. Automated image analysis of protein localization in budding yeast. *Bioinformatics* **23**, i66-71 (2007).
4. Ma, L. et al. A plausible model for the digital response of p53 to DNA damage. *Proc Natl Acad Sci U S A* **102**, 14266-71 (2005).

# Supplementary Materials

- Supplementary Data
- Supplementary Methods
- Supplementary Table 1. Number of expected and actual hits to TF PWMs
- Supplementary Table 2. PFAM DNA-binding domain families with hits to prokaryotes and eukaryotes
- Supplementary Table 3. Average information content from other data sources
- Supplementary Table 4. The number of TF copies per cell